\documentclass{article}
\usepackage{spconf,amsmath,graphicx,hyperref}

\usepackage{cite}
\usepackage{amsmath,amssymb,amsfonts}
\usepackage{algorithmic}
\usepackage{graphicx}
\usepackage{textcomp}
\usepackage{xcolor}
\usepackage{amssymb}
\usepackage{color}
\usepackage{booktabs} 
\usepackage{url}

\usepackage{comment}

\usepackage{booktabs}   % For nicer horizontal lines
\usepackage{adjustbox}  % For \resizebox
\usepackage[table]{xcolor}
\usepackage{hyperref}
\usepackage[capitalize]{cleveref}
\usepackage{url}

\title{Evaluating Objective Speech Quality Metrics for Neural Audio Codecs
}

\name{Luca A. Lanzendörfer \qquad Florian Grötschla}
\address{ETH Zurich}

\begin{document}
%\ninept
%
\maketitle
\begin{abstract}
Neural audio codecs have gained recent popularity for their use in generative modeling as they offer high-fidelity audio reconstruction at low bitrates. While human listening studies remain the gold standard for assessing perceptual quality, they are time-consuming and impractical. In this work, we examine the reliability of existing objective quality metrics in assessing the performance of recent neural audio codecs. To this end, we conduct a MUSHRA listening test on high-fidelity speech signals and analyze the correlation between subjective scores and widely used objective metrics. Our results show that, while some metrics align well with human perception, others struggle to capture relevant distortions. Our findings provide practical guidance for selecting appropriate evaluation metrics when using neural audio codecs for speech.
\end{abstract}


\section{Introduction} \label{sec:intro}

Quality assessment of audio signals is crucial for many use-cases, such as broadcast and streaming, telephony, or generative applications (e.g., text-to-speech models). These applications ideally work with high compression during the generation or transmission of audio content while maintaining high perceptual quality. 
Neural Audio Codecs (NACs) present a promising solution to this compression-quality trade-off, achieving remarkable compression efficiency while maintaining high perceptual audio quality~\cite{kim2024neural}.

Operating below 8 kbps, and in the case of speech as low as 1 kbps~\cite{liu2024semanticodec,siuzdak2024snac,defossez2024moshi}, NACs leverage deep learning architectures to synthesize high-fidelity audio from compressed representations, significantly outperforming traditional audio codecs in reconstruction fidelity at similar bitrates~\cite{siuzdak2024snac,zeghidour2021soundstream,defossez2022high,kumar2024high}. 
However, evaluating the perceptual quality of reconstructed speech signals from these novel neural codecs remains an open question, especially for stereophonic content, where spatial fidelity plays a crucial role in the overall listening experience.

While subjective listening tests such as MUSHRA (MUltiple Stimuli with Hidden Reference and Anchor)~\cite{series2014method} and the related Basic Audio Quality (BAQ) scale, remain the gold standard for evaluating perceptual audio quality, they are costly, time-consuming, and impractical for large-scale evaluations or iterative development. Although listening tests in controlled environments are the most reliable method for assessing BAQ, there is a clear need for reliable, objective evaluation metrics.

Traditional objective quality metrics such as PESQ~\cite{itu2001pesq} and PEAQ~\cite{colomes1999perceptual} are widely used for speech and general audio coding, but their effectiveness in capturing distortions introduced by recent NACs remains underexplored. 

Previous work conducted a comprehensive evaluation of objective perceptual audio quality measures across different application domains, specifically focusing on traditional (non-neural) audio coding and source separation~\cite{torcoli2021objective}.
The study revealed that most objective measures exhibit domain dependence, performing well only in their intended application domain. Therefore, the question remains: How do reconstructed speech signals from recent NACs correlate with widely used objective quality metrics? To this end, we build on previous work~\cite{torcoli2024odaq,torcoli2021objective} and perform a MUSHRA listening test to find the correlation between four current state-of-the-art NACs~\cite{siuzdak2024snac,defossez2024moshi,defossez2022high,kumar2024high} capable of encoding both speech and general audio in a low-bitrate domain. Additionally, we test against two state-of-the-art vocoders that were trained to reconstruct audio from the latent space of the EnCodec model~\cite{siuzdakvocos,san2023discrete}. While this list of models is by no means exhaustive, the selected models span the sub-10 kbps range and have been widely used in practice for audio, speech, and music generation tasks, making them ideal candidates for our study. For our experiments, we use clean speech samples and speech samples containing background audio from ODAQ~\cite{torcoli2024odaq}. 

\begin{figure*}[t!]
    \centering
    \includegraphics[width=\linewidth]{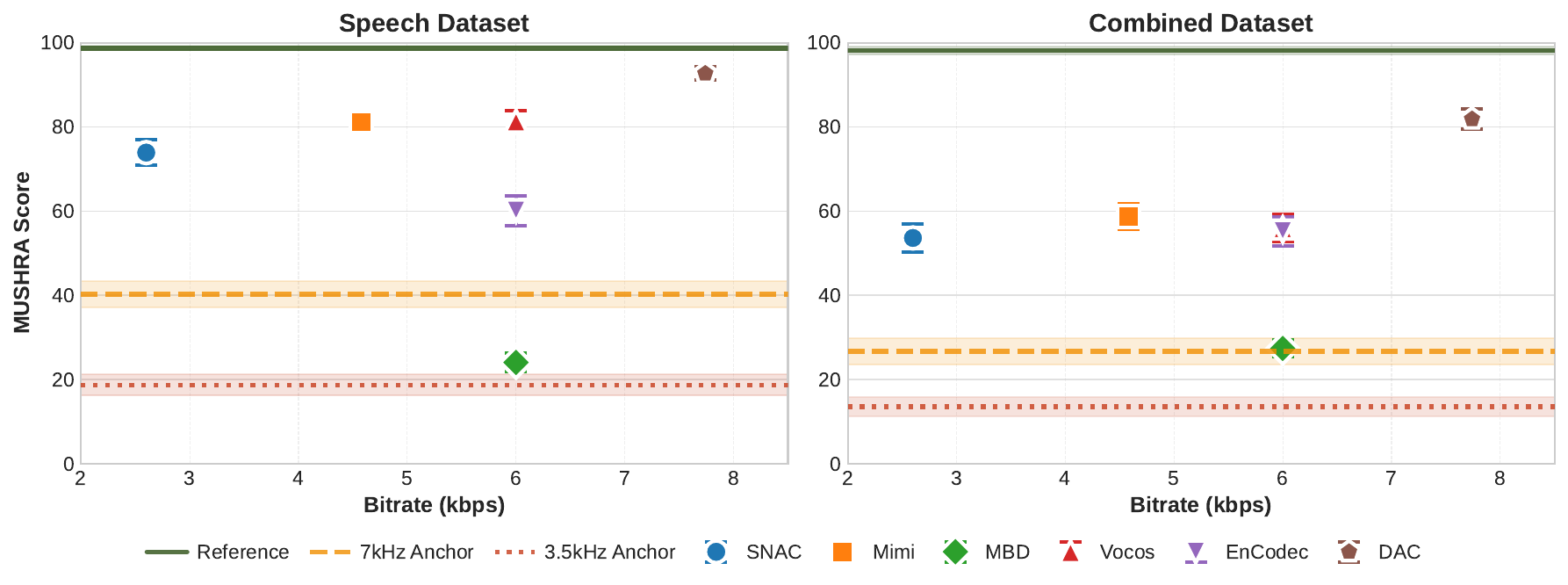}
    \caption{MUSHRA listening test results for speech-only (left) and combined audio (right). The shaded area and whiskers represent 95\% CI. The reference, 7 kHz anchor, and 3.5 kHz anchor were rated at $98.7\pm0.5$, $40.3\pm3.2$, $18.7\pm2.5$ for speech, and $98.1\pm1.0$, $26.7\pm3.2$, $13.6\pm2.2$ for combined audio, respectively.}
    \label{fig:mushra}
\end{figure*}

\noindent Our contributions can be summarized as follows:
\begin{itemize}
\item We review widely used objective quality metrics for audio and speech by conducting a MUSHRA listening study on high-fidelity speech and audio samples encoded and decoded by current state-of-the-art neural audio codecs.
\item We analyze the correlations between the objective quality metrics and the MUSHRA results, giving recommendations on which metrics best capture human perception. This guideline helps future evaluations of novel neural audio codecs for speech and audio. 

\end{itemize}

\section{Methodology}
\label{sec:method}

To evaluate whether widely used objective metrics correlate well with subjective human judgment on high-fidelity speech signals, we run the following experiment. We use the 11 clean speech samples and the 11 combined (speech and background) samples from ODAQ~\cite{torcoli2024odaq}, which are either encoded at 44.1 kHz or 48 kHz. We resample all signals to 48 kHz. As all of the tested NACs do not natively support stereo, we encode and decode each channel separately. We encode each sample with six different models (cf.~\cref{subsec:models}) and conduct a MUSHRA human listening study (cf.~\cref{subsec:mushra}). The resulting scores are then used to calculate the correlation with the objective metrics (cf.~\cref{subsec:metrics}).

\subsection{Models}
\label{subsec:models}

We evaluate six recent neural audio codecs and vocoder architectures for speech and general audio. For EnCodec~\cite{defossez2022high} we use the 24 kHz, 6 kbps model. Multi-Band Diffusion (MBD)~\cite{san2023discrete} is a vocoder that decodes EnCodec latents with band-wise diffusion. Vocos\cite{siuzdakvocos} is another vocoder operating in the Fourier domain that reconstructs EnCodec latents. DAC~\cite{kumar2024high} is able to reconstruct audio at 44.1 kHz, achieving 7.74 kbps and outperforming EnCodec on objective metrics and listening tests across speech, music, and environmental sounds. SNAC~\cite{siuzdak2024snac} extends DAC, achieving between 1-3 kbps bitrate. For our study, we use the 44.1 kHz, 2.6 kbps model. Mimi~\cite{defossez2024moshi}, built on EnCodec, produces acoustic and semantic tokens, compressing 24 kHz speech to 4.4 kbps. The selected models by no means cover the entire spectrum of available NACs, especially since there has been a large body of recent work on neural speech codecs. However, compared to neural speech codecs, the selected models are widely used for all types of audio (general audio, speech, and music) and also span the sub-10 kbps range, making them ideal candidates to evaluate both speech-only and combined audio.

\begin{table*}[]
\small
    \centering
    \caption{Correlations between objective quality metrics and MUSHRA scores for speech-only audio.}

\begin{tabular}{lllllllll}
\toprule
Metric & DAC~\cite{kumar2024high} & EnCodec~\cite{defossez2022high} & MBD~\cite{san2023discrete} & Mimi~\cite{defossez2024moshi} & SNAC~\cite{siuzdak2024snac} & Vocos~\cite{siuzdakvocos} & Pearson $\rho$ & Kendall $\tau$ \\
\midrule
SCOREQ & \cellcolor[rgb]{0.5364705882352944,0.7411764705882354,0.5294117647058824} 0.11 & \cellcolor[rgb]{0.8491349480968859,0.9169550173010381,0.6755709342560555} 0.34 & \cellcolor[rgb]{0.9568627450980393,0.6392156862745099,0.5403921568627452} 0.88 & \cellcolor[rgb]{0.7461284121491736,0.8594386774317571,0.6007996924259902} 0.25 & \cellcolor[rgb]{0.8127797001153404,0.8966551326412918,0.6491810841983853} 0.31 & \cellcolor[rgb]{0.6203767781622453,0.7885428681276433,0.5554632833525568} 0.17 & 0.937 & 0.769 \\
PESQ & \cellcolor[rgb]{0.5364705882352944,0.7411764705882354,0.5294117647058824} 3.90 & \cellcolor[rgb]{0.9977700884275279,0.9576316801230297,0.7789465590157632} 2.66 & \cellcolor[rgb]{0.9568627450980393,0.6392156862745099,0.5403921568627452} 1.82 & \cellcolor[rgb]{0.8612533640907344,0.9237216455209535,0.6843675509419455} 3.27 & \cellcolor[rgb]{0.9966935793925413,0.9371780084582852,0.7449288735101883} 2.56 & \cellcolor[rgb]{0.9393771626297579,0.9667820069204153,0.7713494809688581} 3.04 & 0.886 & 0.721 \\
STOI & \cellcolor[rgb]{0.5364705882352944,0.7411764705882354,0.5294117647058824} 0.97 & \cellcolor[rgb]{0.8612533640907344,0.9237216455209535,0.6843675509419455} 0.93 & \cellcolor[rgb]{0.9568627450980393,0.6392156862745099,0.5403921568627452} 0.85 & \cellcolor[rgb]{0.6737716262975779,0.8186851211072665,0.5720415224913495} 0.95 & \cellcolor[rgb]{0.9993079584775086,0.9868512110726645,0.8275432525951558} 0.91 & \cellcolor[rgb]{0.806720492118416,0.8932718185313342,0.6447827758554403} 0.94 & 0.885 & 0.636 \\
NORESQA-MOS & \cellcolor[rgb]{0.5364705882352944,0.7411764705882354,0.5294117647058824} 4.46 & \cellcolor[rgb]{0.5669819300269128,0.75840061514802,0.5388850442137639} 4.42 & \cellcolor[rgb]{0.9568627450980393,0.6392156862745099,0.5403921568627452} 3.09 & \cellcolor[rgb]{0.5593540945790082,0.7540945790080739,0.5365167243367935} 4.42 & \cellcolor[rgb]{0.5746097654748175,0.7627066512879662,0.5412533640907344} 4.41 & \cellcolor[rgb]{0.5364705882352944,0.7411764705882354,0.5294117647058824} 4.46 & 0.871 & 0.420 \\
ViSQOL Speech & \cellcolor[rgb]{0.5364705882352944,0.7411764705882354,0.5294117647058824} 4.06 & \cellcolor[rgb]{0.7400692041522492,0.8560553633217993,0.5964013840830451} 3.73 & \cellcolor[rgb]{0.9568627450980393,0.6392156862745099,0.5403921568627452} 2.18 & \cellcolor[rgb]{0.8248981161091888,0.9034217608612073,0.6579777008842753} 3.55 & \cellcolor[rgb]{0.9573394848135334,0.9766243752402922,0.7944790465205691} 3.23 & \cellcolor[rgb]{0.7195386389850059,0.8445213379469435,0.586251441753172} 3.76 & 0.850 & 0.592 \\
WARP-Q raw & \cellcolor[rgb]{0.5364705882352944,0.7411764705882354,0.5294117647058824} 1.60 & \cellcolor[rgb]{0.9983852364475202,0.9693194925028835,0.7983852364475201} 2.01 & \cellcolor[rgb]{0.9568627450980393,0.6392156862745099,0.5403921568627452} 2.32 & \cellcolor[rgb]{0.8127797001153404,0.8966551326412918,0.6491810841983853} 1.78 & \cellcolor[rgb]{0.9951557093425606,0.897439446366782,0.7018685121107267} 2.10 & \cellcolor[rgb]{0.9124336793540947,0.9520184544405998,0.7366551326412919} 1.86 & 0.847 & 0.702 \\
WARP-Q normalized & \cellcolor[rgb]{0.5364705882352944,0.7411764705882354,0.5294117647058824} 0.54 & \cellcolor[rgb]{0.9983852364475202,0.9693194925028835,0.7983852364475201} 0.43 & \cellcolor[rgb]{0.9568627450980393,0.6392156862745099,0.5403921568627452} 0.34 & \cellcolor[rgb]{0.8127797001153404,0.8966551326412918,0.6491810841983853} 0.49 & \cellcolor[rgb]{0.9951557093425606,0.897439446366782,0.7018685121107267} 0.40 & \cellcolor[rgb]{0.9124336793540947,0.9520184544405998,0.7366551326412919} 0.47 & 0.847 & 0.702 \\
2f-Model & \cellcolor[rgb]{0.5364705882352944,0.7411764705882354,0.5294117647058824} 60.28 & \cellcolor[rgb]{0.8370165321030374,0.9101883890811228,0.6667743175701654} 52.53 & \cellcolor[rgb]{0.9568627450980393,0.6392156862745099,0.5403921568627452} 33.05 & \cellcolor[rgb]{0.9994617454825068,0.9897731641676278,0.8324029219530951} 46.05 & \cellcolor[rgb]{0.9393771626297579,0.9667820069204153,0.7713494809688581} 49.09 & \cellcolor[rgb]{0.7764244521337949,0.8763552479815456,0.6227912341407152} 54.39 & 0.836 & 0.542 \\
C-SI-SNR & \cellcolor[rgb]{0.5364705882352944,0.7411764705882354,0.5294117647058824} 11.20 & \cellcolor[rgb]{0.903452518262207,0.9470972702806613,0.7250903498654364} 5.99 & \cellcolor[rgb]{0.9568627450980393,0.6392156862745099,0.5403921568627452} -3.40 & \cellcolor[rgb]{0.7764244521337949,0.8763552479815456,0.6227912341407152} 7.97 & \cellcolor[rgb]{0.9483583237216455,0.9717031910803537,0.7829142637447136} 4.97 & \cellcolor[rgb]{0.9999231064975009,0.9985390234525182,0.8469819300269127} 3.85 & 0.745 & 0.466 \\
SA-SDR & \cellcolor[rgb]{0.5364705882352944,0.7411764705882354,0.5294117647058824} 11.19 & \cellcolor[rgb]{0.903452518262207,0.9470972702806613,0.7250903498654364} 5.98 & \cellcolor[rgb]{0.9568627450980393,0.6392156862745099,0.5403921568627452} -3.43 & \cellcolor[rgb]{0.7764244521337949,0.8763552479815456,0.6227912341407152} 7.96 & \cellcolor[rgb]{0.9483583237216455,0.9717031910803537,0.7829142637447136} 4.96 & \cellcolor[rgb]{0.9999231064975009,0.9985390234525182,0.8469819300269127} 3.82 & 0.744 & 0.462 \\
SI-SDR & \cellcolor[rgb]{0.5364705882352944,0.7411764705882354,0.5294117647058824} 11.19 & \cellcolor[rgb]{0.903452518262207,0.9470972702806613,0.7250903498654364} 5.98 & \cellcolor[rgb]{0.9568627450980393,0.6392156862745099,0.5403921568627452} -3.42 & \cellcolor[rgb]{0.7764244521337949,0.8763552479815456,0.6227912341407152} 7.96 & \cellcolor[rgb]{0.9483583237216455,0.9717031910803537,0.7829142637447136} 4.96 & \cellcolor[rgb]{0.9999231064975009,0.9985390234525182,0.8469819300269127} 3.82 & 0.744 & 0.462 \\
SI-SNR & \cellcolor[rgb]{0.5364705882352944,0.7411764705882354,0.5294117647058824} 11.19 & \cellcolor[rgb]{0.903452518262207,0.9470972702806613,0.7250903498654364} 5.98 & \cellcolor[rgb]{0.9568627450980393,0.6392156862745099,0.5403921568627452} -3.42 & \cellcolor[rgb]{0.7764244521337949,0.8763552479815456,0.6227912341407152} 7.96 & \cellcolor[rgb]{0.9483583237216455,0.9717031910803537,0.7829142637447136} 4.96 & \cellcolor[rgb]{0.9999231064975009,0.9985390234525182,0.8469819300269127} 3.82 & 0.744 & 0.462 \\
SDR & \cellcolor[rgb]{0.5364705882352944,0.7411764705882354,0.5294117647058824} 12.25 & \cellcolor[rgb]{0.9528489042675894,0.9741637831603229,0.7886966551326413} 7.36 & \cellcolor[rgb]{0.9568627450980393,0.6392156862745099,0.5403921568627452} 0.90 & \cellcolor[rgb]{0.782483660130719,0.8797385620915034,0.6271895424836602} 9.69 & \cellcolor[rgb]{0.9988465974625145,0.9780853517877739,0.812964244521338} 6.01 & \cellcolor[rgb]{0.9985390234525182,0.972241445597847,0.8032449058054595} 5.87 & 0.731 & 0.487 \\
SNR & \cellcolor[rgb]{0.5364705882352944,0.7411764705882354,0.5294117647058824} 11.42 & \cellcolor[rgb]{0.9618300653594772,0.9790849673202615,0.8002614379084968} 6.74 & \cellcolor[rgb]{0.9568627450980393,0.6392156862745099,0.5403921568627452} 1.00 & \cellcolor[rgb]{0.8430757400999617,0.9135717031910804,0.6711726259131104} 8.42 & \cellcolor[rgb]{0.9994617454825068,0.9897731641676278,0.8324029219530951} 5.92 & \cellcolor[rgb]{0.9973087274125336,0.9488658208381392,0.7643675509419455} 5.05 & 0.728 & 0.460 \\
ViSQOL Audio & \cellcolor[rgb]{0.5364705882352944,0.7411764705882354,0.5294117647058824} 4.17 & \cellcolor[rgb]{0.9618300653594772,0.9790849673202615,0.8002614379084968} 3.61 & \cellcolor[rgb]{0.9568627450980393,0.6392156862745099,0.5403921568627452} 2.91 & \cellcolor[rgb]{0.9968473663975395,0.9400999615532488,0.7497885428681278} 3.37 & \cellcolor[rgb]{0.7643060361399462,0.8695886197616302,0.6139946174548252} 3.91 & \cellcolor[rgb]{0.9573394848135334,0.9766243752402922,0.7944790465205691} 3.62 & 0.691 & 0.450 \\
DNSMOS & \cellcolor[rgb]{0.5364705882352944,0.7411764705882354,0.5294117647058824} 3.88 & \cellcolor[rgb]{0.9971549404075356,0.9459438677431756,0.7595078815840062} 3.61 & \cellcolor[rgb]{0.9568627450980393,0.6392156862745099,0.5403921568627452} 3.44 & \cellcolor[rgb]{0.6127489427143407,0.7842368319876971,0.5530949634755864} 3.85 & \cellcolor[rgb]{0.806720492118416,0.8932718185313342,0.6447827758554403} 3.77 & \cellcolor[rgb]{0.5364705882352944,0.7411764705882354,0.5294117647058824} 3.88 & 0.683 & 0.503 \\
NISQA & \cellcolor[rgb]{0.5364705882352944,0.7411764705882354,0.5294117647058824} 4.01 & \cellcolor[rgb]{0.9790080738177624,0.7440369088811994,0.5935409457900808} 3.18 & \cellcolor[rgb]{0.9568627450980393,0.6392156862745099,0.5403921568627452} 3.07 & \cellcolor[rgb]{0.5364705882352944,0.7411764705882354,0.5294117647058824} 4.01 & \cellcolor[rgb]{0.8127797001153404,0.8966551326412918,0.6491810841983853} 3.77 & \cellcolor[rgb]{0.6966551326412918,0.831603229527105,0.5791464821222606} 3.88 & 0.563 & 0.419 \\
PEAQ & \cellcolor[rgb]{0.5364705882352944,0.7411764705882354,0.5294117647058824} -2.34 & \cellcolor[rgb]{0.9568627450980393,0.6392156862745099,0.5403921568627452} -3.42 & \cellcolor[rgb]{0.9971549404075356,0.9459438677431756,0.7595078815840062} -3.01 & \cellcolor[rgb]{0.9596309111880047,0.6523183391003461,0.5470357554786621} -3.41 & \cellcolor[rgb]{0.9957708573625529,0.9161399461745483,0.7176163014225297} -3.07 & \cellcolor[rgb]{0.972087658592849,0.711280276816609,0.5769319492502885} -3.34 & 0.145 & 0.122 \\
NORESQA & \cellcolor[rgb]{0.9748558246828143,0.7243829296424451,0.5835755478662054} 6.09 & \cellcolor[rgb]{0.9259054209919262,0.9594002306805075,0.7540023068050751} 5.42 & \cellcolor[rgb]{0.9568627450980393,0.6392156862745099,0.5403921568627452} 6.20 & \cellcolor[rgb]{0.9950019223375625,0.8927643214148405,0.6979315647827758} 5.83 & \cellcolor[rgb]{0.9977700884275279,0.9576316801230297,0.7789465590157632} 5.68 & \cellcolor[rgb]{0.5364705882352944,0.7411764705882354,0.5294117647058824} 4.91 & 0.145 & 0.217 \\
\bottomrule
\end{tabular}
    \label{tab:correlations_speech}
\end{table*}

\subsection{Objective Metrics}
\label{subsec:metrics}

We evaluate audio quality with objective measures spanning distortion, perceptual quality, intelligibility, and no-reference mean opinion score (MOS) estimation. Distortion metrics include SNR~\cite{le2019sdr}, scale-invariant SI-SNR and C-SI-SNR~\cite{luo2018tasnet,ni2021wpd++}, as well as SDR, SI-SDR, and SA-SDR~\cite{1643671,le2019sdr,von2022sa}. Perceptual metrics comprise PESQ~\cite{itu2001pesq} and PEAQ (we report the Basic ODG combining 11 MOVs)~\cite{colomes1999perceptual,torcoli2021objective,kabal2002examination}, in addition to the 2f-Model that aggregates ADB and AvgModDiff1~\cite{kastner2019efficient,torcoli2021objective}; we exclude POLQA due to lack of access and its limited gains over PESQ~\cite{povcta2015subjective,torcoli2021objective}. Speech intelligibility is measured by STOI~\cite{5495701}, and spectrogram similarity by ViSQOL (Audio at 48 kHz and Speech at 16 kHz)~\cite{chinen2020visqol,hines2012speech}. Non-intrusive MOS predictors include DNSMOS~\cite{reddy2021dnsmos} and NISQA~\cite{mittag2021nisqa}, while NORESQA and NORESQA-MOS estimate relative and absolute MOS without exact references~\cite{noresqa,noresqamos}. Furthermore, we report WARP-Q~\cite{warpq} and SCOREQ~\cite{ragano2024scoreq}, two recently proposed metrics.

\subsection{Subjective Listening Test}
\label{subsec:mushra}

We conducted a double-blind MUSHRA~\cite{series2014method} listening test to evaluate the perceptual quality of neural audio codecs. We recruited participants from a crowd-sourcing MUSHRA tool.\footnote{\url{https://www.mabyduck.com}} Stimuli were drawn from ODAQ~\cite{torcoli2024odaq}, resampled to 48 kHz, and loudness normalized to -28 dB to avoid clipping. 
Participants were required to take short breaks between trials to avoid fatigue. Each trial consisted of nine stimuli: The signals reconstructed from the models as well as a hidden reference and two anchors (downsampled reference to 3.5 kHz and 7 kHz, respectively). Listeners rated each stimulus on a 0–100 scale relative to the reference. Stimuli orders were randomized and playback was unrestricted.
In accordance with the MUSHRA protocol, we removed the raters who scored the reference below 90 at least twice (out of 11 trials, this results in a failure rate of 18\% and therefore above the 15\% recommended by the protocol). After filtering, we were left with 11 participants for the speech-only test and 17 participants for the combined audio test.

\subsection{Correlation Analysis}
To assess the relationship between objective quality metrics and human perception, we compute Pearson’s correlation coefficient ($\rho$) and Kendall’s tau ($\tau$) between each metric and the mean MUSHRA scores across all participants.
Pearson’s correlation coefficient is defined as:
\begin{equation}
\rho = \frac{\sum (x_i - \bar{x})(y_i - \bar{y})}{\sqrt{\sum (x_i - \bar{x})^2} \sqrt{\sum (y_i - \bar{y})^2}}
\end{equation}
where $x_i$ and $y_i$ are the individual values of the mean MUSHRA scores and the corresponding objective metric for model/sound snippet combination $i$, and $\bar{x}$ and $\bar{y}$ are their respective means. Pearson’s correlation measures the strength of the linear relationship between two variables, making it useful for assessing direct proportionality.
Kendall’s tau is computed as:
\begin{equation}
    \tau = \frac{C - D}{\frac{1}{2}n(n-1)}
\end{equation}
where $C$ is the number of concordant pairs, $D$ is the number of discordant pairs among all pairs, and $n$ is the total number of combinations.
Higher values of $\rho$ and $\tau$ indicate stronger alignment between the objective metric and human perception. 
Pearson and Kendall correlation coefficients provide different insights into the relationship between objective metrics and MUSHRA scores. Pearson’s correlation measures the strength of the linear relationship between two variables. However, it is sensitive to outliers and assumes a normally distributed dataset. Kendall’s correlation, on the other hand, is based on ranking and measures the ordinal association between two variables. This makes it more robust to non-linear relationships and less sensitive to outliers.

\begin{table*}[]
\small
    \centering

\begin{tabular}{lllllllll}

\toprule
Metric & DAC~\cite{kumar2024high} & EnCodec~\cite{defossez2022high} & MBD~\cite{san2023discrete} & Mimi~\cite{defossez2024moshi} & SNAC~\cite{siuzdak2024snac} & Vocos~\cite{siuzdakvocos} & Pearson $\rho$ & Kendall $\tau$ \\
\midrule
PESQ & \cellcolor[rgb]{0.5364705882352944,0.7411764705882354,0.5294117647058824} 3.61 & \cellcolor[rgb]{0.9982314494425221,0.9663975394079201,0.7935255670895809} 2.44 & \cellcolor[rgb]{0.9568627450980393,0.6392156862745099,0.5403921568627452} 1.57 & \cellcolor[rgb]{0.9982314494425221,0.9663975394079201,0.7935255670895809} 2.45 & \cellcolor[rgb]{0.9922337562475971,0.8086120722798923,0.6270665128796618} 1.91 & \cellcolor[rgb]{0.9946943483275663,0.8834140715109573,0.6900576701268744} 2.14 & 0.903 & 0.696 \\
SCOREQ & \cellcolor[rgb]{0.5364705882352944,0.7411764705882354,0.5294117647058824} 0.17 & \cellcolor[rgb]{0.7946020761245676,0.8865051903114187,0.6359861591695504} 0.28 & \cellcolor[rgb]{0.9568627450980393,0.6392156862745099,0.5403921568627452} 0.64 & \cellcolor[rgb]{0.800661284121492,0.8898885044213765,0.6403844675124952} 0.28 & \cellcolor[rgb]{0.7946020761245676,0.8865051903114187,0.6359861591695504} 0.28 & \cellcolor[rgb]{0.7946020761245676,0.8865051903114187,0.6359861591695504} 0.28 & 0.869 & 0.506 \\
2f-Model & \cellcolor[rgb]{0.5364705882352944,0.7411764705882354,0.5294117647058824} 49.61 & \cellcolor[rgb]{0.6890272971933872,0.8272971933871588,0.5767781622452903} 46.64 & \cellcolor[rgb]{0.9568627450980393,0.6392156862745099,0.5403921568627452} 27.19 & \cellcolor[rgb]{0.9968473663975395,0.9400999615532488,0.7497885428681278} 35.39 & \cellcolor[rgb]{0.9079430988081508,0.9495578623606306,0.7308727412533642} 41.37 & \cellcolor[rgb]{0.8248981161091888,0.9034217608612073,0.6579777008842753} 43.57 & 0.816 & 0.399 \\
STOI & \cellcolor[rgb]{0.5364705882352944,0.7411764705882354,0.5294117647058824} 0.90 & \cellcolor[rgb]{0.8309573241061131,0.906805074971165,0.6623760092272204} 0.83 & \cellcolor[rgb]{0.9568627450980393,0.6392156862745099,0.5403921568627452} 0.65 & \cellcolor[rgb]{0.8989619377162631,0.9446366782006921,0.7193079584775088} 0.81 & \cellcolor[rgb]{0.9954632833525567,0.9067896962706652,0.7097424067666281} 0.73 & \cellcolor[rgb]{0.9528489042675894,0.9741637831603229,0.7886966551326413} 0.79 & 0.772 & 0.577 \\
SNR & \cellcolor[rgb]{0.5364705882352944,0.7411764705882354,0.5294117647058824} 9.35 & \cellcolor[rgb]{0.9393771626297579,0.9667820069204153,0.7713494809688581} 5.54 & \cellcolor[rgb]{0.9568627450980393,0.6392156862745099,0.5403921568627452} 0.11 & \cellcolor[rgb]{0.9483583237216455,0.9717031910803537,0.7829142637447136} 5.44 & \cellcolor[rgb]{0.9982314494425221,0.9663975394079201,0.7935255670895809} 4.04 & \cellcolor[rgb]{0.9966935793925413,0.9371780084582852,0.7449288735101883} 3.40 & 0.765 & 0.511 \\
SDR & \cellcolor[rgb]{0.5364705882352944,0.7411764705882354,0.5294117647058824} 9.93 & \cellcolor[rgb]{0.9259054209919262,0.9594002306805075,0.7540023068050751} 5.49 & \cellcolor[rgb]{0.9568627450980393,0.6392156862745099,0.5403921568627452} -1.41 & \cellcolor[rgb]{0.8944713571703192,0.9421760861207229,0.7135255670895811} 5.99 & \cellcolor[rgb]{0.9983852364475202,0.9693194925028835,0.7983852364475201} 3.48 & \cellcolor[rgb]{0.9979238754325259,0.960553633217993,0.7838062283737024} 3.26 & 0.754 & 0.534 \\
SA-SDR & \cellcolor[rgb]{0.5364705882352944,0.7411764705882354,0.5294117647058824} 8.95 & \cellcolor[rgb]{0.8430757400999617,0.9135717031910804,0.6711726259131104} 4.43 & \cellcolor[rgb]{0.9568627450980393,0.6392156862745099,0.5403921568627452} -6.59 & \cellcolor[rgb]{0.8491349480968859,0.9169550173010381,0.6755709342560555} 4.33 & \cellcolor[rgb]{0.9483583237216455,0.9717031910803537,0.7829142637447136} 2.34 & \cellcolor[rgb]{0.9932641291810842,0.9963091118800461,0.8407381776239907} 1.34 & 0.727 & 0.512 \\
SI-SDR & \cellcolor[rgb]{0.5364705882352944,0.7411764705882354,0.5294117647058824} 9.00 & \cellcolor[rgb]{0.8430757400999617,0.9135717031910804,0.6711726259131104} 4.43 & \cellcolor[rgb]{0.9568627450980393,0.6392156862745099,0.5403921568627452} -6.64 & \cellcolor[rgb]{0.8491349480968859,0.9169550173010381,0.6755709342560555} 4.33 & \cellcolor[rgb]{0.9483583237216455,0.9717031910803537,0.7829142637447136} 2.34 & \cellcolor[rgb]{0.9932641291810842,0.9963091118800461,0.8407381776239907} 1.34 & 0.727 & 0.513 \\
SI-SNR & \cellcolor[rgb]{0.5364705882352944,0.7411764705882354,0.5294117647058824} 9.00 & \cellcolor[rgb]{0.8430757400999617,0.9135717031910804,0.6711726259131104} 4.43 & \cellcolor[rgb]{0.9568627450980393,0.6392156862745099,0.5403921568627452} -6.64 & \cellcolor[rgb]{0.8491349480968859,0.9169550173010381,0.6755709342560555} 4.33 & \cellcolor[rgb]{0.9483583237216455,0.9717031910803537,0.7829142637447136} 2.34 & \cellcolor[rgb]{0.9932641291810842,0.9963091118800461,0.8407381776239907} 1.34 & 0.727 & 0.512 \\
C-SI-SNR & \cellcolor[rgb]{0.5364705882352944,0.7411764705882354,0.5294117647058824} 9.07 & \cellcolor[rgb]{0.8370165321030374,0.9101883890811228,0.6667743175701654} 4.47 & \cellcolor[rgb]{0.9568627450980393,0.6392156862745099,0.5403921568627452} -7.20 & \cellcolor[rgb]{0.8430757400999617,0.9135717031910804,0.6711726259131104} 4.43 & \cellcolor[rgb]{0.934886582083814,0.964321414840446,0.7655670895809306} 2.43 & \cellcolor[rgb]{0.9797923875432527,0.9889273356401385,0.8233910034602077} 1.42 & 0.709 & 0.508 \\
ViSQOL Audio & \cellcolor[rgb]{0.5364705882352944,0.7411764705882354,0.5294117647058824} 3.76 & \cellcolor[rgb]{0.9753018069973086,0.9864667435601692,0.8176086120722799} 2.98 & \cellcolor[rgb]{0.9568627450980393,0.6392156862745099,0.5403921568627452} 2.09 & \cellcolor[rgb]{0.9963860053825452,0.9313341022683583,0.7352095347943098} 2.67 & \cellcolor[rgb]{0.7195386389850059,0.8445213379469435,0.586251441753172} 3.49 & \cellcolor[rgb]{0.9994617454825068,0.9897731641676278,0.8324029219530951} 2.89 & 0.644 & 0.232 \\
WARP-Q normalized & \cellcolor[rgb]{0.5364705882352944,0.7411764705882354,0.5294117647058824} 0.44 & \cellcolor[rgb]{0.9842829680891965,0.9913879277201076,0.8291733948481353} 0.35 & \cellcolor[rgb]{0.9845444059976932,0.7702422145328719,0.6068281430219147} 0.28 & \cellcolor[rgb]{0.9990003844675125,0.9810073048827375,0.8178239138792771} 0.34 & \cellcolor[rgb]{0.9568627450980393,0.6392156862745099,0.5403921568627452} 0.26 & \cellcolor[rgb]{0.9942329873125721,0.8693886966551327,0.678246828143022} 0.30 & 0.632 & 0.486 \\
WARP-Q raw & \cellcolor[rgb]{0.5364705882352944,0.7411764705882354,0.5294117647058824} 1.97 & \cellcolor[rgb]{0.9842829680891965,0.9913879277201076,0.8291733948481353} 2.27 & \cellcolor[rgb]{0.9845444059976932,0.7702422145328719,0.6068281430219147} 2.51 & \cellcolor[rgb]{0.9990003844675125,0.9810073048827375,0.8178239138792771} 2.31 & \cellcolor[rgb]{0.9568627450980393,0.6392156862745099,0.5403921568627452} 2.59 & \cellcolor[rgb]{0.9942329873125721,0.8693886966551327,0.678246828143022} 2.43 & 0.632 & 0.485 \\
ViSQOL Speech & \cellcolor[rgb]{0.5364705882352944,0.7411764705882354,0.5294117647058824} 3.59 & \cellcolor[rgb]{0.8612533640907344,0.9237216455209535,0.6843675509419455} 3.04 & \cellcolor[rgb]{0.9568627450980393,0.6392156862745099,0.5403921568627452} 1.79 & \cellcolor[rgb]{0.9988465974625145,0.9780853517877739,0.812964244521338} 2.60 & \cellcolor[rgb]{0.9886966551326413,0.7898961937716263,0.61679354094579} 2.06 & \cellcolor[rgb]{0.9999231064975009,0.9985390234525182,0.8469819300269127} 2.68 & 0.593 & 0.434 \\
PEAQ & \cellcolor[rgb]{0.5364705882352944,0.7411764705882354,0.5294117647058824} -2.39 & \cellcolor[rgb]{0.9734717416378317,0.7178316032295271,0.580253748558247} -3.44 & \cellcolor[rgb]{0.9900807381776241,0.7964475201845447,0.6201153402537487} -3.35 & \cellcolor[rgb]{0.9568627450980393,0.6392156862745099,0.5403921568627452} -3.53 & \cellcolor[rgb]{0.9976163014225298,0.9547097270280661,0.7740868896578238} -3.08 & \cellcolor[rgb]{0.9665513264129182,0.6850749711649367,0.5636447520184544} -3.48 & 0.582 & -0.045 \\
NORESQA & \cellcolor[rgb]{0.5364705882352944,0.7411764705882354,0.5294117647058824} 4.95 & \cellcolor[rgb]{0.5974932718185315,0.7756247597078048,0.5483583237216456} 5.02 & \cellcolor[rgb]{0.9568627450980393,0.6392156862745099,0.5403921568627452} 6.42 & \cellcolor[rgb]{0.7119108035371012,0.8402153018069973,0.5838831218762015} 5.17 & \cellcolor[rgb]{0.6661437908496733,0.8143790849673203,0.569673202614379} 5.12 & \cellcolor[rgb]{0.6661437908496733,0.8143790849673203,0.569673202614379} 5.12 & 0.374 & 0.219 \\
NISQA & \cellcolor[rgb]{0.9982314494425221,0.9663975394079201,0.7935255670895809} 1.91 & \cellcolor[rgb]{0.903452518262207,0.9470972702806613,0.7250903498654364} 2.06 & \cellcolor[rgb]{0.9568627450980393,0.6392156862745099,0.5403921568627452} 1.63 & \cellcolor[rgb]{0.6585159554017685,0.8100730488273742,0.5673048827374088} 2.23 & \cellcolor[rgb]{0.9966935793925413,0.9371780084582852,0.7449288735101883} 1.87 & \cellcolor[rgb]{0.5364705882352944,0.7411764705882354,0.5294117647058824} 2.30 & 0.193 & 0.205 \\
NORESQA-MOS & \cellcolor[rgb]{0.5364705882352944,0.7411764705882354,0.5294117647058824} 4.00 & \cellcolor[rgb]{0.6051211072664362,0.7799307958477509,0.550726643598616} 3.99 & \cellcolor[rgb]{0.9859284890426759,0.7767935409457901,0.6101499423298731} 3.83 & \cellcolor[rgb]{0.9693194925028835,0.6981776239907728,0.5702883506343714} 3.82 & \cellcolor[rgb]{0.6737716262975779,0.8186851211072665,0.5720415224913495} 3.97 & \cellcolor[rgb]{0.9568627450980393,0.6392156862745099,0.5403921568627452} 3.81 & 0.124 & 0.017 \\
DNSMOS & \cellcolor[rgb]{0.9979238754325259,0.960553633217993,0.7838062283737024} 2.66 & \cellcolor[rgb]{0.9637831603229527,0.6719723183391003,0.5570011534025374} 2.59 & \cellcolor[rgb]{0.6966551326412918,0.831603229527105,0.5791464821222606} 2.75 & \cellcolor[rgb]{0.5364705882352944,0.7411764705882354,0.5294117647058824} 2.77 & \cellcolor[rgb]{0.9568627450980393,0.6392156862745099,0.5403921568627452} 2.59 & \cellcolor[rgb]{0.7764244521337949,0.8763552479815456,0.6227912341407152} 2.73 & -0.073 & 0.058 \\
\bottomrule
\end{tabular}
\caption{Correlations between objective quality metrics and MUSHRA scores for combined (speech and background) audio.}
    \label{tab:correlations}
\end{table*}

\section{Results}
\label{sec:results}

\subsection{Subjective Results}
\label{subsec:subjective}

The results from the MUSHRA listening study (cf.~\cref{fig:mushra}) reveal several findings across speech-only audio and combined audio (speech with background audio). DAC~\cite{kumar2024high} consistently outperform other NACs, but given its higher bitrate, this is not entirely surprising. Interestingly, for the speech-only test, experienced participants were sometimes unable to distinguish the reference signal from DAC, which underlines the excellent perceptual quality of this model. 
Additionally, it is worth noting that the results on combined audio showed generally lower performance compared to speech-only conditions, likely due to the compression artifacts being more noticeable in the broader frequency spectrum of mixed content, whereas speech-only audio occupies a more limited bandwidth. 
Furthermore, we observe a substantial difference in performance between Vocos~\cite{siuzdakvocos} and EnCodec~\cite{defossez2022high} when comparing speech-only and combined audio results. While Vocos significantly outperformed EnCodec in the speech-only setting, it performed comparably to EnCodec in the combined audio setting. This discrepancy can likely be attributed to Vocos being trained predominantly on clean speech data, making it less robust to mixed-content scenarios. Additionally, MBD~\cite{san2023discrete} performed significantly worse than expected, likely due to the presence of audible spatial artifacts.

\subsection{Correlation Analysis}
\label{subsec:correlation}

The correlation analysis reveals several findings on the predictive performance of existing objective quality metrics as can be seen in \cref{tab:correlations_speech} for speech-only and in \cref{tab:correlations} for combined audio. 
Surprisingly, PESQ a model proposed more than two decades ago exhibited among the highest correlation with subjective ratings (Pearson $\rho$ = 0.903 for combined audio and $\rho$ = 0.886 for speech-only audio), underscoring its effectiveness as a predictor of perceptual quality, and outperforming more recent approaches. SCOREQ, a recent metric demonstrated the strongest correlation on speech-only audio ($\rho$ = 0.937), while still performing excellently in the combined audio setting ($\rho$ = 0.869). 
For the combined audio, the correlation of ViSQOL-Speech was notably lower ($\rho$ = 0.593) compared to its performance on speech-only data ($\rho$ = 0.85), indicating that it struggles to capture degradations outside of the domain for which it was designed. Similarly, PEAQ, DNSMOS, NISQA, and ViSQOL-Audio exhibited weak correlations in both settings, indicating that they are not suitable for the evaluation of NACs. 
It is also noteworthy to mention that while the 2f-Model is based on the same intermediate scores that PEAQ uses to compute its final output, the 2f-Model consistently outperforms PEAQ. 
Overall, PESQ and SCOREQ stand out as the models with the highest correlations and can thus be regarded as a good practical choice.
Statistical significance tests (p-values $<<$ 0.05) confirmed the reliability of these observed differences and provide a strong validation of our findings.

\section{Conclusion}

In this work, we evaluated commonly-used objective quality metrics on neural audio codecs. We found that PESQ, one of the earliest metrics, demonstrated strong correlation across both clean speech and combined audio, making it a reliable choice for practical quality assessment of neural audio codecs together with SCOREQ, a recently proposed metric. While some more recent metrics (e.g., WARP-Q, NORESQA-MOS) showed strong performance on speech-only content, these metrics struggled to capture distortions introduced by neural audio codecs, particularly in audio scenes with background sounds. Furthermore, it is interesting to note that simpler methods (e.g., SNR and SDR) seem to be performing similarly to some metrics specifically designed to measure human perception, sometimes even outperforming them. 
We believe these findings provide practical guidance for selecting appropriate evaluation metrics when developing and using neural audio codecs for speech data.

\ninept
\bibliographystyle{IEEEbib}
\bibliography{strings,refs}

\begin{thebibliography}{10}

\bibitem{kim2024neural}
Minje Kim and Jan Skoglund,
\newblock ``Neural speech and audio coding,''
\newblock {\em arXiv preprint arXiv:2408.06954}, 2024.

\bibitem{liu2024semanticodec}
Haohe Liu et~al.,
\newblock ``Semanticodec: An ultra low bitrate semantic audio codec for general sound,''
\newblock {\em arXiv preprint arXiv:2405.00233}, 2024.

\bibitem{siuzdak2024snac}
Hubert Siuzdak et~al.,
\newblock ``Snac: Multi-scale neural audio codec,''
\newblock {\em arXiv preprint arXiv:2410.14411}, 2024.

\bibitem{defossez2024moshi}
Alexandre D{\'e}fossez et~al.,
\newblock ``Moshi: a speech-text foundation model for real-time dialogue,''
\newblock {\em arXiv preprint arXiv:2410.00037}, 2024.

\bibitem{zeghidour2021soundstream}
Neil Zeghidour et~al.,
\newblock ``Soundstream: An end-to-end neural audio codec,''
\newblock {\em IEEE/ACM Transactions on Audio, Speech, and Language Processing}, vol. 30, pp. 495--507, 2021.

\bibitem{defossez2022high}
Alexandre D{\'e}fossez et~al.,
\newblock ``High fidelity neural audio compression,''
\newblock {\em arXiv preprint arXiv:2210.13438}, 2022.

\bibitem{kumar2024high}
Rithesh Kumar et~al.,
\newblock ``High-fidelity audio compression with improved rvqgan,''
\newblock {\em Advances in Neural Information Processing Systems}, vol. 36, 2024.

\bibitem{series2014method}
B~Series,
\newblock ``Method for the subjective assessment of intermediate quality level of audio systems,''
\newblock {\em International Telecommunication Union Radiocommunication Assembly}, vol. 2, 2014.

\bibitem{itu2001pesq}
{International Telecommunication Union},
\newblock ``Perceptual evaluation of speech quality (pesq): An objective method for end-to-end speech quality assessment of narrow-band telephone networks and speech codecs,''
\newblock ITU-T Recommendation P.862, International Telecommunication Union, Geneva, Switzerland, 02 2001,
\newblock Approved under WTSA Resolution 1 procedure on 23 February 2001.

\bibitem{colomes1999perceptual}
Catherine Colomes et~al.,
\newblock ``Perceptual quality assessment for digital audio: Peaq-the new itu standard for objective measurement of the perceived audio quality,''
\newblock in {\em Audio Engineering Society Conference: 17th International Conference: High-Quality Audio Coding}. Audio Engineering Society, 1999.

\bibitem{torcoli2021objective}
Matteo Torcoli et~al.,
\newblock ``Objective measures of perceptual audio quality reviewed: An evaluation of their application domain dependence,''
\newblock {\em IEEE/ACM Transactions on Audio, Speech, and Language Processing}, vol. 29, pp. 1530--1541, 2021.

\bibitem{torcoli2024odaq}
Matteo Torcoli et~al.,
\newblock ``Odaq: Open dataset of audio quality,''
\newblock in {\em ICASSP 2024-2024 IEEE International Conference on Acoustics, Speech and Signal Processing (ICASSP)}. IEEE, 2024, pp. 836--840.

\bibitem{siuzdakvocos}
Hubert Siuzdak,
\newblock ``Vocos: Closing the gap between time-domain and fourier-based neural vocoders for high-quality audio synthesis,''
\newblock in {\em The Twelfth International Conference on Learning Representations}.

\bibitem{san2023discrete}
Robin San~Roman et~al.,
\newblock ``From discrete tokens to high-fidelity audio using multi-band diffusion,''
\newblock {\em Advances in neural information processing systems}, vol. 36, pp. 1526--1538, 2023.

\bibitem{le2019sdr}
Jonathan Le~Roux et~al.,
\newblock ``Sdr--half-baked or well done?,''
\newblock in {\em ICASSP 2019-2019 IEEE International Conference on Acoustics, Speech and Signal Processing (ICASSP)}. IEEE, 2019, pp. 626--630.

\bibitem{luo2018tasnet}
Yi~Luo and Nima Mesgarani,
\newblock ``Tasnet: time-domain audio separation network for real-time, single-channel speech separation,''
\newblock in {\em 2018 IEEE International Conference on Acoustics, Speech and Signal Processing (ICASSP)}. IEEE, 2018, pp. 696--700.

\bibitem{ni2021wpd++}
Zhaoheng Ni et~al.,
\newblock ``Wpd++: An improved neural beamformer for simultaneous speech separation and dereverberation,''
\newblock in {\em 2021 IEEE Spoken Language Technology Workshop (SLT)}. IEEE, 2021, pp. 817--824.

\bibitem{1643671}
E.~Vincent et~al.,
\newblock ``Performance measurement in blind audio source separation,''
\newblock {\em IEEE Transactions on Audio, Speech, and Language Processing}, vol. 14, no. 4, pp. 1462--1469, 2006.

\bibitem{von2022sa}
Thilo von Neumann et~al.,
\newblock ``Sa-sdr: A novel loss function for separation of meeting style data,''
\newblock in {\em ICASSP 2022-2022 IEEE International Conference on Acoustics, Speech and Signal Processing (ICASSP)}. IEEE, 2022, pp. 6022--6026.

\bibitem{kabal2002examination}
Peter Kabal et~al.,
\newblock ``An examination and interpretation of itu-r bs. 1387: Perceptual evaluation of audio quality,''
\newblock {\em TSP Lab Technical Report, Dept. Electrical \& Computer Engineering, McGill University}, pp. 1--89, 2002.

\bibitem{kastner2019efficient}
Thorsten Kastner and J{\"u}rgen Herre,
\newblock ``An efficient model for estimating subjective quality of separated audio source signals,''
\newblock in {\em 2019 IEEE Workshop on Applications of Signal Processing to Audio and Acoustics (WASPAA)}. IEEE, 2019, pp. 95--99.

\bibitem{povcta2015subjective}
Peter Po{\v{c}}ta and John~G Beerends,
\newblock ``Subjective and objective assessment of perceived audio quality of current digital audio broadcasting systems and web-casting applications,''
\newblock {\em IEEE Transactions on broadcasting}, vol. 61, no. 3, pp. 407--415, 2015.

\bibitem{5495701}
Cees~H. Taal et~al.,
\newblock ``A short-time objective intelligibility measure for time-frequency weighted noisy speech,''
\newblock in {\em 2010 IEEE International Conference on Acoustics, Speech and Signal Processing}, 2010, pp. 4214--4217.

\bibitem{chinen2020visqol}
Michael Chinen et~al.,
\newblock ``Visqol v3: An open source production ready objective speech and audio metric,''
\newblock in {\em 2020 twelfth international conference on quality of multimedia experience (QoMEX)}. IEEE, 2020, pp. 1--6.

\bibitem{hines2012speech}
Andrew Hines and Naomi Harte,
\newblock ``Speech intelligibility prediction using a neurogram similarity index measure,''
\newblock {\em Speech Communication}, vol. 54, no. 2, pp. 306--320, 2012.

\bibitem{reddy2021dnsmos}
Chandan~KA Reddy et~al.,
\newblock ``Dnsmos: A non-intrusive perceptual objective speech quality metric to evaluate noise suppressors,''
\newblock in {\em ICASSP 2021-2021 IEEE International Conference on Acoustics, Speech and Signal Processing (ICASSP)}. IEEE, 2021, pp. 6493--6497.

\bibitem{mittag2021nisqa}
Gabriel Mittag et~al.,
\newblock ``Nisqa: A deep cnn-self-attention model for multidimensional speech quality prediction with crowdsourced datasets,''
\newblock {\em arXiv preprint arXiv:2104.09494}, 2021.

\bibitem{noresqa}
Pranay Manocha et~al.,
\newblock ``{NORESQA}: A framework for speech quality assessment using non-matching references,''
\newblock in {\em Thirty-Fifth Conference on Neural Information Processing Systems}, 2021.

\bibitem{noresqamos}
Pranay Manocha and Anurag Kumar,
\newblock ``Speech quality assessment through mos using non-matching references,''
\newblock in {\em Interspeech}, 2022.

\bibitem{warpq}
Wissam~A. Jassim et~al.,
\newblock ``Speech quality assessment with warp-q: From similarity to subsequence dynamic time warp cost,''
\newblock {\em IET Signal Processing}, vol. 16, no. 9, pp. 1050--1070, 2022.

\bibitem{ragano2024scoreq}
Alessandro Ragano et~al.,
\newblock ``Scoreq: Speech quality assessment with contrastive regression,''
\newblock in {\em Advances in Neural Information Processing Systems}, A.~Globerson, L.~Mackey, D.~Belgrave, A.~Fan, U.~Paquet, J.~Tomczak, and C.~Zhang, Eds. 2024, vol.~37, pp. 105702--105729, Curran Associates, Inc.

\end{thebibliography}

\end{document}